# Optimisation and artifacts of photothermal excitation of microresonators

*Liping Kevin Ge[1,2], Alessandro Tuniz[2,3], Martijn de Sterke[2,3], James M. Zavislan[4], Thomas G. Brown[4], Sascha Martin[5] and David Martinez-Martin[\*][1,2]*

[1] School of Biomedical Engineering, Faculty of Engineering and IT, University of Sydney, NSW, 2006 Australia
[2] The University of Sydney Nano Institute, University of Sydney, NSW, 2006, Australia
[3] Institute of Photonics and Optical Science (IPOS), School of Physics, University of Sydney, Sydney, NSW, 2006, Australia
[4] The Institute of Optics, University of Rochester, Rochester, New York, 14627, USA
[5] Department of Physics, University of Basel, Basel, 4056, Switzerland

[\*]*Correspondence:* David Martinez-Martin: Email | david.martinezmartin@sydney.edu.au

## ABSTRACT

The excitation of microresonators using focused intensity modulated light, known as photothermal excitation, is gaining significant attention due to its capacity to accurately excite microresonators without distortions, even in liquid environments, which is driving key advancements in atomic force microscopy and related technologies. Despite progress in the development of coatings, the conversion of light into mechanical movement remains largely inefficient, limiting resonator movements to tens of nanometres even when milliwatts of optical power are used. Moreover, how photothermal efficiency depends on the relative position of a microresonator along the propagation axis of the photothermal beam remains poorly studied, hampering our understanding of the conversion of light into mechanical motion. Here, we perform photothermal measurements in air and water using cantilever microresonators and a custom-built picobalance, and determine how photothermal efficiency changes along the propagation beam axis. We identify that far out-of-band laser emission can lead to visual misidentification of the beam waist, resulting in a drop of photothermal efficiency of up to one order of magnitude. Our measurements also unveil that the beam waist is not always the position of highest photothermal efficiency, and can reduce the efficiency up to 20% for silicon cantilevers with trapezoidal cross section.

## INTRODUCTION

Recent years have been marked by a renewed interest in microelectromechanical systems (MEMS), atomic force microscopy (AFM), and picobalance devices, which enable nanosensing applications across biomedicine, nanotechnology and materials science [1-4]. Such applications typically require the accurate excitation of oscillations in microcantilevers and microresonators in vacuum, air, or liquids. Amongst them, liquids present the most challenging environment, because their added mass and viscosity reduce the quality factor of microcantilevers below 10, thus diminishing the overall measurement accuracy.

Many approaches can be used to excite cantilever oscillations: these include *acoustic, magnetic, electrostatic, and Lorentz-force-induced* excitations. Of these, *acoustic* approaches are the most common, wherein oscillations are driven by a vibrating dither piezoelectric within the cantilever holder [5, 6]. However, these methods often produce spurious resonance peaks associated with the excitation of other parts of the larger instrument, interfering with the cantilever movement. This is particularly detrimental in liquids, where the cantilever quality factor is low (<10) [5], thereby demanding high driving powers. *Magnetic* excitation instead require a magnetic bead attached to the cantilever or a magnetic functionalisation of the cantilever itself, which can degrade over time and can be toxic to biological samples [7, 8]. In addition, the magnetic coil near the cantilever precludes the simultaneous use of a transmission optical microscope, limiting practicality. *Electrostatic* excitation requires the application of an alternating bias voltage between the cantilever and an electrode, which induces a detrimental surface charge diffusion when working in liquids [9]. Finally, *Lorentz-force-induced* excitation requires alternating current through a cantilever under a static magnetic field, but is limited to very specific cantilever geometries that can transport electric currents, significantly increasing the cantilever's temperature, thus hindering biological applications [10].

All the above limitations can be overcome by exciting the cantilever with an intensity-modulated light beam, via the *photothermal excitation method* [11]. Although this method was developed before the invention of AFMs, and was used to excite and detect resonator oscillations via a beam deflection scheme [11], it has recently become available in commercial AFMs. With this scheme, an intensity-modulated light beam is positioned near the fixed end of the cantilever, and induces a localised temperature gradient which dilates a portion of the cantilever at the modulation frequency, producing a well defined cantilever movement [12, 13]. While the first designs used metal-coated cantilevers [14], uncoated cantilevers have recently been developed [4]. Most importantly, regardless of whether the cantilever is in vacuum, air or liquid, this excitation method is accurately described by a simple damped harmonic oscillator response [15], and works over a wide frequency bandwidth of more than 10 MHz [16]. However, large cantilever oscillation amplitudes (> 10 nm) with a low average optical power (< 100 µW) remain challenging, particularly when the cantilevers are in liquid [16-19].

Delicate samples such as live mammalian cells or yeast [4, 20] benefit from using low optical powers to prevent damage, which leads to small cantilever amplitudes. However, large cantilever amplitudes (>10 nm) are known to increase the quality of AFM measurements when quantifying mass [4, 20], rheology [21], and force-distance AFM based measurements [22]. Larger cantilever amplitudes operating at lower powers can be achieved by increasing the photothermal efficiency, which has recently become an increasingly active field of research: approaches typically involve either physical or chemical modifications of cantilevers (e.g., by partially coating them with light-absorbing photo-acoustic materials such as gold, carbon or nanoparticle coating layers), or the use of different cantilever geometries. For example, coatings can enhance optical energy conversion into mechanical motion, increasing a cantilever's oscillation amplitude by up to 6 times for a given power [23-25], whereas trapezoidal cantilever cross-sections have been reported to perform better than their rectangular counterparts [26], noting that the location of the exciting laser spot on the cantilever also plays an important role [15]. However, the dependence of the photothermal excitation efficiency relative to the location of the cantilever along the propagation axis of the beam is yet to be characterised in detail (Fig. 1a).

Here we present a detailed characterisation of the changes in cantilever excitation efficiency along the propagation axis (here: z-axis) of a photothermal excitation beam using a picobalance [4, 20], which is mounted on an inverted optical microscope in similar fashion to commercially available bio-AFMs (illustrated in the Fig. 1a schematic). Our measurements reveal that, due to a long-wavelength artifact resulting from far out-of-band laser emission, the longitudinal position that leads to the highest photothermal excitation efficiency can easily be misjudged when using an inverted optical microscope, which may reduce the photothermal efficiency by up to approximately one order of magnitude, even though the error introduced in the laser working distance is approximately 5%. We unambiguously identify that the origin of this artifact is due to spurious laser light, and we provide convenient and practical measures to address it. Our work also unveils that the longitudinal position that provides the highest photothermal excitation efficiency is not always at the beam waist but depends on the cantilever properties and can reduce the photothermal efficiency by up to 20%. Considering that photothermal excitation is becoming essential in many AFMs [22] and the emerging picobalance technology [4], often used in conjunction with an inverted optical microscope, we expect this work to inform optimal design procedures and operation guidelines for related technologies.

## RESULTS and DISCUSSIONS

### Photothermal excitation efficiency along the beam propagation axis

AFMs typically include a laser (wavelength 852 nm in our setup) to determine the cantilever movement. It comes with two lateral degrees of freedom (x- and y) to adjust the position of the laser on the cantilever, whereas its longitudinal (z) position is generally fixed. Photothermal excitation relies on the introduction of a second, intensity modulated laser (wavelength: 405 nm in our setup) [27] whose position is adjusted in a similar fashion. Figure 1b shows a zoomed-in schematic of the experimental setup used in this work: it consists of a custom-built inertial picobalance, a technology whose principal inventor is one of us (DMM) and which is used to characterise the mass of living mammalian and yeast cells, as well as mechanical and rheological cell properties

[4, 20, 21, 28-30]. The picobalance is mounted on an inverted microscope, which can simultaneously provide sample information from both transmitted differential interference contrast microscopy and fluorescence microscopy (Fig. 1a-b). Note that this setup can also operate as an AFM by adding a piezo scanner [31]. However, unlike conventional AFMs, the position of each laser in the picobalance can be adjusted both laterally and longitudinally via piezo-motor positioners, enabling us to characterise the photothermal efficiency along the propagation axis of the photothermal beam (Fig. 1c).

As a first experiment, we recorded cantilever resonance curves (amplitude and phase versus the excitation frequency) driven with photothermal excitation for different relative positions of the cantilever along the longitudinal axis of the photothermal excitation beam. Figure 2a shows the amplitude and phase vs frequency of a cantilever in water at two different longitudinal positions. The red curve in this figure corresponds to the longitudinal position that provides the highest photothermal efficiency (optimal position), while the blue curve corresponds to the longitudinal position at which the photothermal laser beam waist is located on the cantilever according to the optical microscope. Figure 2b shows measurements of the cantilever amplitude at resonance (highest amplitude in a resonance curve) for different relative longitudinal positions of the cantilever. In all these measurements the excitation beam strikes the cantilever near its base at the optimal lateral location [26, 32]. Once this location was found, it was kept constant during the experiments, and only the longitudinal position was changed. The origin of the z axis was chosen at the optimal position. The longitudinal position labelled "apparent beam waist" corresponds to the position at which the beam waist of the photothermal beam coincides with the cantilever, according to optical images. The measurements were performed with average optical powers of the photothermal beam varying from 25.9 µW to 46.3 µW (Fig. 2b). Optical images of the photothermal beam at the cantilever plane were recorded for different longitudinal positions. Fig 2c depicts an optical image (top image) of the beam at the cantilever plane for the longitudinal position that provided the highest cantilever amplitude; the lower image shows the optical image of the beam for the longitudinal position at which the beam waist appears at the cantilever plane.

These measurements were performed in both air and liquid environments and for cantilevers with standard geometries (rectangular shape with rectangular cross section, triangular shape with rectangular cross section, and rectangular shape with trapezoidal cross section cantilevers) and made of common materials, including gold coated silicon nitride and silicon cantilevers, and bare silicon cantilevers without coating (Fig. S1). A summary of the results is shown in Fig 2d, where the vertical axis represents the enhancement in efficiency, which is the ratio between the resonance amplitude at the optimal longitudinal position for a given cantilever type and that at the apparent beam waist position. One would expect the highest photothermal efficiency occur near the beam waist of the photothermal beam; however, our experiments showed that the cantilever oscillation amplitude, and therefore the photothermal efficiency, increased by 2-8 times at a longitudinal position that was far away from the beam waist as identified by using the optical microscope (Fig. 2b, d and Fig. S1). Surprisingly, as depicted in Figures 2b and S1, the z position for which the photothermal efficiency was the highest depends on the cantilever type and is at least at 2450 μm below the discerned beam waist (here termed "apparent beam waist"), therefore well outside the 206 μm confocal range of our laser (Fig 1c). Yet, it is important to note that by design, the working distance of our laser is 46.2 mm; hence, a change in the working distance of 2450 μm represents only a 5% change. We now investigate and discuss the origin of this discrepancy in detail.

**Identifying the true beam waist position**

The photothermal laser beam in our picobalance is a Gaussian beam with a waist (i.e., minimum 1/e intensity radius) of 3.65 μm (Fig 1c). Therefore, if we aim the photothermal beam on a cantilever and we move the cantilever in the longitudinal direction (we do this by approaching or withdrawing the laser source to or from the cantilever), the optical power measured beneath the cantilever should be minimum when the cantilever is at the beam waist, as at that point the cantilever would screen most of the laser light (Fig 3a-c). Gold-coated cantilevers are very opaque to the wavelength of our photothermal laser (405 nm), and in silicon cantilevers the penetration depth of this wavelength is only ~100 nm [33]. Thus, the cantilevers are generally very opaque at the excitation wavelength.

Figure 3a shows a schematic of the experimental set-up to perform such optical power measurements beneath the cantilever. The photothermal beam is aligned in the lateral directions and can move longitudinally. Below the cantilever an optical power sensor (Thorlabs S120VC) measures the residual optical power of the photothermal beam that is not screened by the cantilever (Fig. 3b). Results are shown in Fig. 3d, where the blue curve corresponds to the cantilever amplitude at resonance for different z positions of the beam relative to the cantilever, and the red curve shows simultaneous measurements of the residual (unscreened) optical power below the cantilever. As before, the origin of the z axis is chosen to be the longitudinal position that provides the highest resonant cantilever amplitude for a given photothermal beam power. These data demonstrate that the beam waist position identified with the optical microscope (apparent beam waist in Fig. 3c) differs from the real position of the beam waist (beam waist in Fig 3c) by approximately 2450 μm.

We note that Gaussian beams exhibit a *focal shift*, which causes the position of the beam waist to differ from that of the nominal focal point of a lens (Fig. 3d) [34]. For our Gaussian beam is calculated to be 32 μm, which is consistent with our observations. Moreover, the wavelength of 405 nm of our photothermal beam is at the edge of the visible part of the spectrum, whereas the objective (CFI Plan Fluor 10X from Nikon) in our setup is aberration corrected at 550 nm. Our measurements (Fig. S2) indicate a position shift of only 50 μm at the 405 nm wavelength. Therefore neither the focal shift, nor chromatic aberration can explain the observed 2450 μm discrepancy between the real and the apparent beam waists.

**Origin of the Apparent Beam Waist**

With the focal shift and chromatic aberration ruled out, we considered whether the apparent beam waist could be linked to the wavelength distribution of the laser source. Although more than 99% of the laser light produced by a laser diode is usually centred around the nominal wavelength, there is still some light produced outside that range. In order to be able to acquire fluorescent and/or differential interference contrast images, optical filters are placed in front of the camera (Fig 1a): a longpass filter (FEL0450, Thorlabs, US) with a cut-off wavelength of 450 nm to prevent the photothermal beam from saturating the images and a short-pass filter (FES0750, Thorlabs, US) with a cut-off wavelength of 750 nm to prevent the read-out laser from saturating the images. Therefore, the apparent beam waist detected could be the beam waist of unfiltered radiation at a wavelength above 450 nm rather than the actual beam waist of the attenuated 405 nm radiation. To investigate this, we mounted an additional and identical longpass filter to see whether the observed radiation would attenuate further. However, we did not observe further attenuation nor further displacement of the apparent beam waist, implying that the observed beam waist with the microscope was that of wavelengths higher than 450 nm. To confirm this, we returned to the initial configuration of filters (Fig. 1a) and mounted a bandpass filter (15117 Edmund Optics, US) at the output of the 405 nm laser and next to the existing neutral density filter. The bandpass filter was centred at wavelength of 405 nm and had a bandwidth of 10 nm. With this new filter in place, the apparent beam waist was not detectable by the microscope in the presence of the longpass filter, confirming that it corresponded to a wavelength above 450 nm.

We then removed the longpass filter whilst keeping the 405 nm bandpass filter and repeated the photothermal efficiency measurements simultaneously with optical power measurements beneath the cantilever (similar to Fig. 3c). We found that under these conditions the apparent beam waist and the real beam waist essentially coincide with an error of approximately 300 μm (Fig. S3). This error compares to the 206 μm confocal parameter of the laser; the 50 μm chromatic aberration of the objective at 405 nm; and the 32 μm Gaussian focal shift. Since without the longpass filter the laser light prevents the acquisition of optical information from the sample, it should only be removed for the purpose of laser alignment. These experimental results demonstrate that the laser diode generates radiation well outside the nominal range. Thus, although generally more than 99% of optical power is within a few nanometres of the nominal wavelength, there is still a portion of the optical power at wavelengths that are well away from this wavelength, which may interfere with measurements of the position of the beam waist (Fig. 4a) or impact the sample.

We then repeated the measurements of Fig. 2 but with the bandpass filter in place (Figure S4). The results demonstrate that for silicon nitride cantilevers with rectangular cross section, the highest photothermal efficiency takes place near the beam waist and within the confocal range of the laser. However, for silicon cantilevers with trapezoidal cross section, we identified that the highest photothermal efficiency takes place at a longitudinal position outside the confocal range of the laser, increasing the photothermal efficiency by up to 20% compare to that at the beam waist (Figure S4). Thus more accurate models are required to better understand photothermal excitation under different conditions.

Finally, we conducted a spectral analysis of our laser source to characterise its power distribution within the wavelength range of 350–850 nm. Figure 4b show the results of these measurements with the long-pass filter (red curve) and without (blue curve). The blue curve confirms that approximately 99.9% of the laser emission is within a window of 10 nm centred at 405 nm. However, the red curves shows that the laser still emits approximately 0.1% of the power at wavelengths hundreds of nm away from the intended 405 nm. Furthermore, the red spectrum also demonstrates that when the 450 nm long-pass filter is used, the 405 nm is essentially fully attenuated and non-intended radiation with wavelengths of 540 nm and 610 nm become dominant, giving rise to the artifact of the apparent beam waist found with the optical microscope.

## CONCLUSION

In conclusion, we have characterised the changes in photothermal excitation efficiency along the propagation axis of a photothermal beam. We have found that the beam waist location can be erroneously identified from visual inspection alone, which reduces the photothermal efficiency by nearly an order of magnitude. This surprising result is caused by spurious low-power long-wavelength laser radiation that is well above the specified (narrow) range. This radiation dominates in the presence of filters commonly used in correlative atomic force and optical microscopies to prevent camera saturation and which enable the acquisition of fluorescence and DIC from the sample. We have provided an effective means to manage this phenomenon in order to optimise the photothermal excitation of cantilevers and resonators. Moreover, our results demonstrate that the highest photothermal efficiency occurs at different longitudinal positions for different cantilever types. In particular we identified that silicon nitride cantilevers with rectangular cross section are most efficiently excited near the laser beam waist within the confocal range of the laser, whereas silicon cantilevers with trapezoidal cross section are best excited outside this confocal range. These results point to the need to adjust the relative longitudinal position of a microcantilever or resonator to optimise the photothermal efficiency. This is particularly important when very low laser powers ($\ll 1$ mW) are required, as for instance when working with live cells [4, 20], but also to optimise photothermal excitation when using different types of cantilevers, and cantilever chip thicknesses, as these can vary by hundreds of micrometres, affecting the longitudinal position of the cantilever relative to the photothermal beam. Given the increasing use and applications of photothermal excitation in multiple systems such as AFMs and picobalances, we expect this work to significantly improve the performance of present and future instruments and to inform the design of new devices.

## METHODS

### Experimental setup

A custom-built picobalance device is shown in Fig. 1a-b. It includes two lasers, a blue-violet photothermal excitation laser with wavelength 405 nm (Thorlabs, US), and an infrared detection laser with wavelength 852 nm (Schäfter + Kirchhoff GmbH, Germany). The 405 nm laser is driven in Current Control Mode by a laser diode controller (LDC500, Stanford Research System, US). To improve the stability of this laser, its temperature is kept constant with a Peltier element controlled by a thermoelectric controller (LDC500). The 852 nm laser, combined with a four-quadrant Si PIN photodiode (S5980 Hamamatsu, Japan) was used to measure the deflection of the cantilever. A hard-coated bandpass filter (Edmund Optics, US) with a central wavelength of 850 nm and 25 nm bandwidth with an optical density of 4 is located in front of the photodiode. A Ti2-E inverted optical microscopy (Nikon, Japan) mounted with a CFI Plan Fluor 10X (Nikon, Japan) objective can capture real-time differential interference contrast (DIC) images during system operation. ECS nano-positioners (Attocube, Germany) controlled by AMC100 controllers (Attocube, Germany) can accurately adjust both lasers' lateral position relative to the cantilever as well as its longitudinal position.

### Beam waist location and Photothermal efficiency

To confirm whether the excitation laser's beam waist observed from the optical microscope coincided with the actual position, we laterally adjusted the excitation laser spot on the cantilever in air for each measurement until a maximum amplitude was obrained. After that, the blue laser was adjusted longitudinally by a ECS nano-positioner with pre-defined increments of 300 μm. The cantilever's oscillation amplitude was registered using the detection laser followed by a photodiode detector and a lock-in amplifier. Two different setups of lock-in amplifiers were used, a CX device from Nanosurf AG (Switzerland) and a BP4.5 control system with an OC5 oscillation controller from Nanonis (Germany). A laser power meter PM100D (Thorlabs, US) was placed underneath the cantilever to measure laser power around the cantilever. The infrared detection laser is switched off to avoid the interference of laser powers with excitation laser during the measurement of transmitted power **(Fig. 1c)**. The excitation laser's longitudinal 'position's origin was set at the position with the highest amplitude. The cantilever oscillation amplitude was estimated from thermal noise measurements using 'Sader's method' to calibrate the cantilever's spring constant [35-37].

**Determination of oscillation amplitude along laser propagation axis**

The comprehensive analysis of peak oscillation amplitude for each position along the excitation laser propagation axis was measured for the different types of cantilevers in both air and water. The cantilever is immersed in the water environment within a 35 mm Petri dish (Ibidi, Germany). Two silicon nitride cantilevers were used: Nunano QUEST R 500 TL (rectangular shape with rectangular cross section) and Bruker NP-010D (triangular shape with rectangular cross section). Both cantilevers were coated with the backside of gold. The silicon cantilever MikroMasch NSC35/CR-AU-C with gold-coated (rectangular shape with trapezoidal cross section) and MikroMasch NSC35/NO AL-A without gold coating (rectangular shape with trapezoidal cross section) were also tested for photothermal efficiency enhancement. The detailed parameters and specs of cantilevers used in the experiments are shown in **Table 1**. The input power of the excitation laser is range from 25.9 $\mu W$ to 46.3 $\mu W$ without BP filter and 18.5 $\mu W$ to 33.1 $\mu W$ with BP filter.

**Determination of chromatic aberration of objective**

To investigate if the differences between apparent focus and the beam waist were caused by chromatic aberration from the microscopy objective, we placed a pinhole with 2 $\mu m$ diameter (Edmund Optics) in the center of the 405 nm laser beam and the 550 nm of light emitted from microscopy. We first aligned the laser at the middle of the field of view shown on the microscopy to align the laser and the pinhole on the same axis before placing the pinhole. This avoided the difficulty of alignment without viewing the position. The pinhole was placed on a 35 mm Petri dish (Ibidi) followed by fine-tuning the pinhole position to control the microscopy stage (Nanosurf), ensuring that the laser and the pinhole coincide. The objective height adjusts to find the minimum spot for both 405 nm laser and 550 nm light as the focal position, and the chromatic aberration of the objective is given as the different heights of focal positions.

**Laser power spectrum acquisition**

A reflective collimator directs the source beam to an inverted microscope (Nikon Instruments Eclipse). The light is focused to a mirror using a microscope objective (Olympus 50X), and its reflection is directed to a calibrated imaging spectrometer (IsoPlane SCT 320 – Princeton Instruments) with visible camera (PIXIS). Wavelength dependent intensity spectra between 350-850nm are obtained by binning the counts at each wavelength. The results are shown in Fig. 4(b). Each spectrum, with- and without- the low pass filter, is obtained after maximizing the power density in the central region of the camera following a small focal adjustment, as per the photothermal excitation measurements shown in the schematic of Fig. 3(d).

# Author contributions

DMM scoped the initial project which was discussed with LKG, MdS and AT who provided insightful suggestions. DMM and LKG mounted the picobalance instruments. DMM, LKG and SM implemented required modifications in the picobalances. LKG performed the photothermal efficiency measurements with different cantilevers in air and water. JMZ and TGB provided insightful discussions and suggested the pinhole experiments to measure chromatic aberration, which were performed by LKG and DMM. AT performed the laser power spectrum experiments with support by LKG and DMM. All the authors reviewed and discussed the data. All authors contributed to write the manuscript.

# Acknowledgement

This work has benefited from the SOAR award and Engineering Research Scholarship (ERS) provided by The University of Sydney. We acknowledged D. Stenger, P. McCarthy and the mechanical workshop from The Faculty of Engineering (The University of Sydney) for their help of manufacturing the enclosures that host the picobalances; A. Tonin and the electronic workshop of the physics department (University of Basel) for helping with the beam deflection electronics; the mechanical workshop of the physics department (University of Basel) for their help with the development and manufacturing of the devices; P. Buchmann and P. Argast from the electronic and mechanical workshop of the department of biosystems science and engineering (ETH Zurich) for helping with the temperature controlled system of the enclosures; Nanosurf AG for their technical support customising software, mechanical and electronic components.

# Competing of interests

# Figures

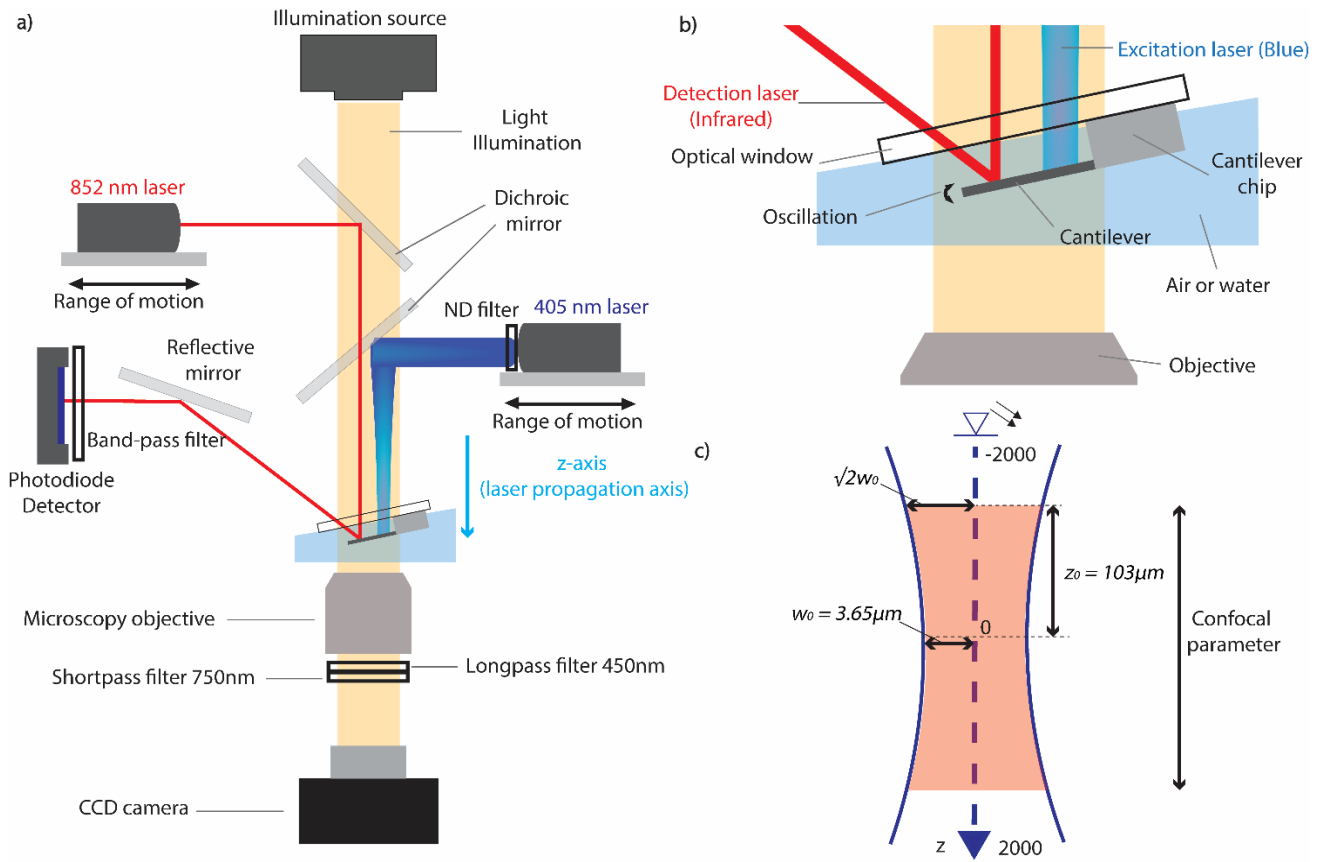

**Figure 1**. **Schematic of the picobalance and Gaussian beam. a)** The picobalance comprises a photothermal excitation laser beam (405 nm laser drawn in blue), which is used to excite the cantilever. The position of each laser can be adjusted independently in three orthogonal directions using piezo-based positioners. The picobalance is mounted on and compatible with an inverted optical microscope. **b)** Detail from a) showing the cantilever and the Gaussian beams reaching the cantilever. The cantilever can be in either air or in water. The microscope objective is underneath the cantilever. **c)** Gaussian beam propagating downwards in the z-direction. The main Gaussian beam parameters are indicated with their actual values for the photothermal beam. The beam waist $w_0 = 3.65$ μm, where the beam radius is smallest. The Rayleigh range $z_0 = 103$ μm corresponds to the propagation distance where the cross section of the beam is twice that at the beam waist. The confocal parameter is twice the Rayleigh range.

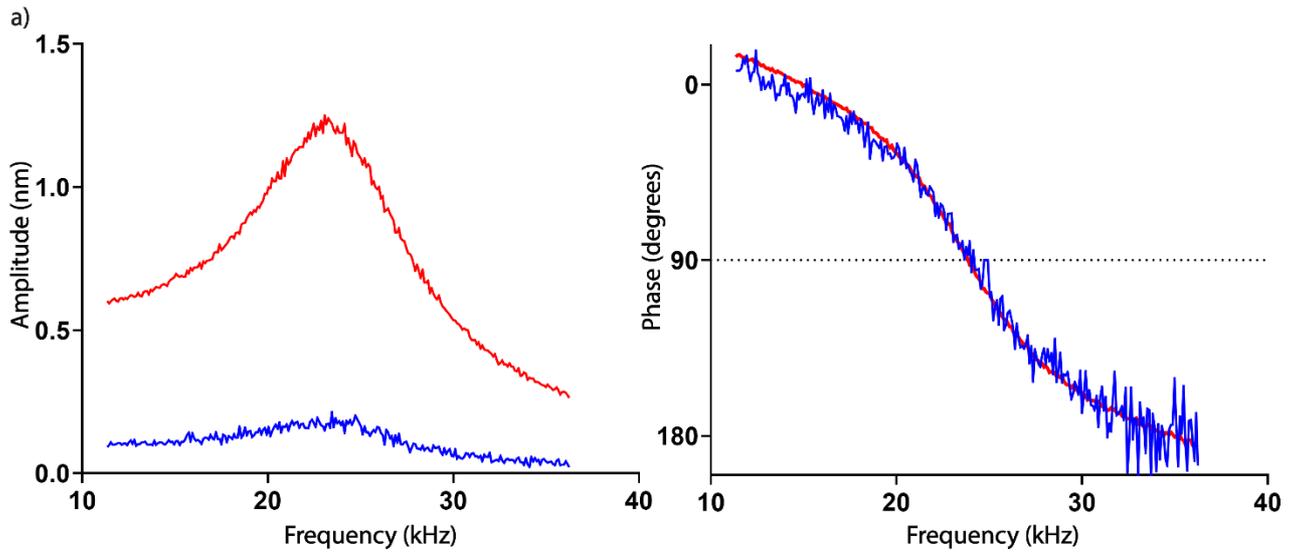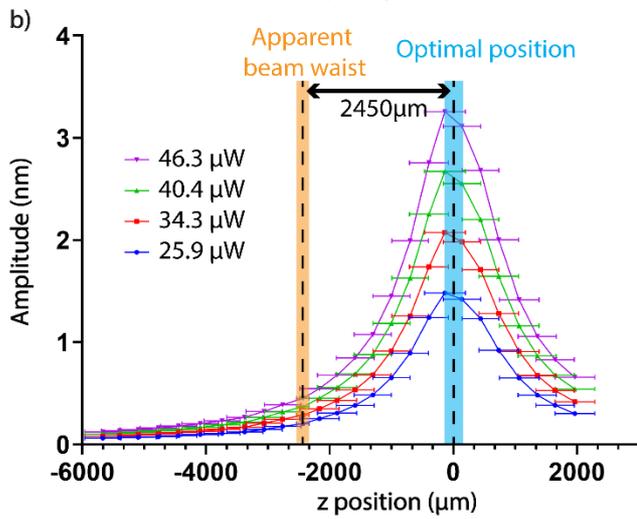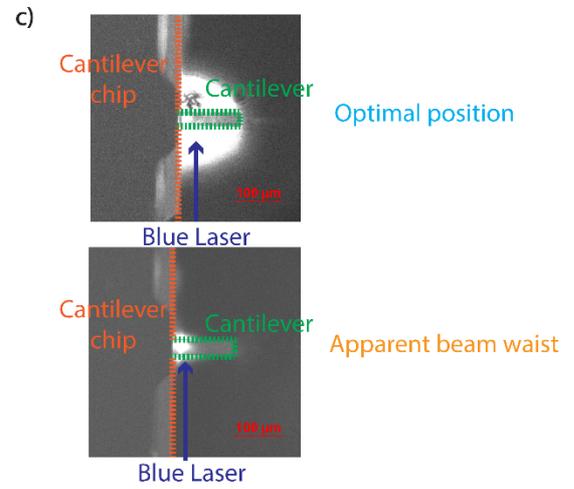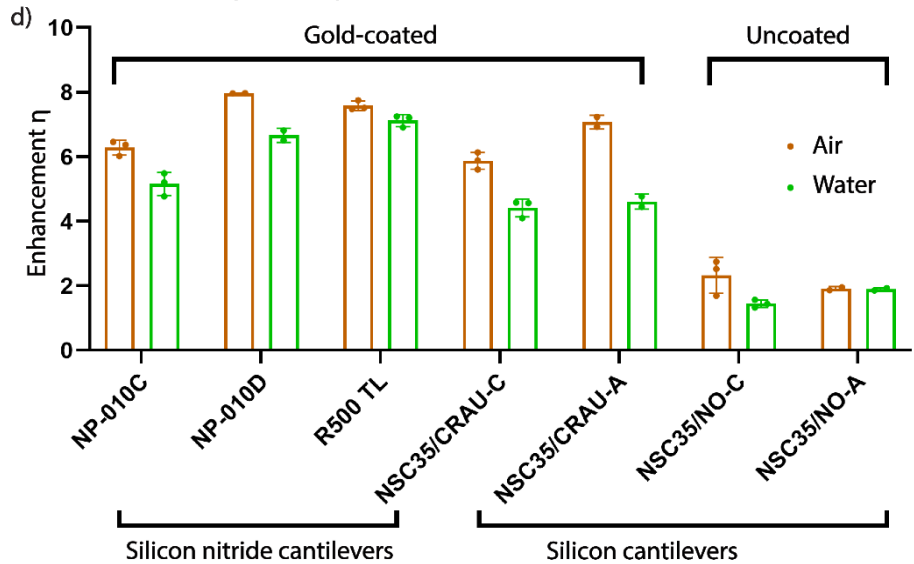

**Figure 2. Photothermal efficiency as a function of the relative longitudinal position of the cantilever. a)** Amplitude and phase response of a cantilever versus frequency for two different cantilever positions. The blue lines correspond to resonance curves (amplitude and phase) of the cantilever when the cantilever is located at the beam waist according to the optical microscope. The red lines show equivalent information as the blue lines but for the optimal position along the propagation axis of the photothermal beam. The measurements are performed in water with a R500 TL cantilever. **b)** Cantilever amplitude at resonance for different relative longitudinal cantilever positions for different optical powers of the photothermal beam. The origin of the axis is chosen to be at the position of maximum (optimal) photothermal efficiency, which is indicated by the dashed line (shaded in blue). The other dashed line (shaded in orange) indicates the longitudinal position at which the cantilever is at the apparent beam waist. The measurements are performed in water with a R500 TL cantilever. **c)** The images show, top: optical image of a cantilever and photothermal beam at the optimal position; bottom: optical image of the cantilever at the apparent beam waist. **d)** Results for similar measurements as in b), but for different cantilever types in both air and water environments. The enhancement factors of the resonance amplitude at the optimal longitudinal position compared to that at the apparent beam waist are shown. All results were conducted for at least two independent cantilever probes and measurements for each probe were repeated at least twice. For each laser position five consecutive resonance curves were acquired.

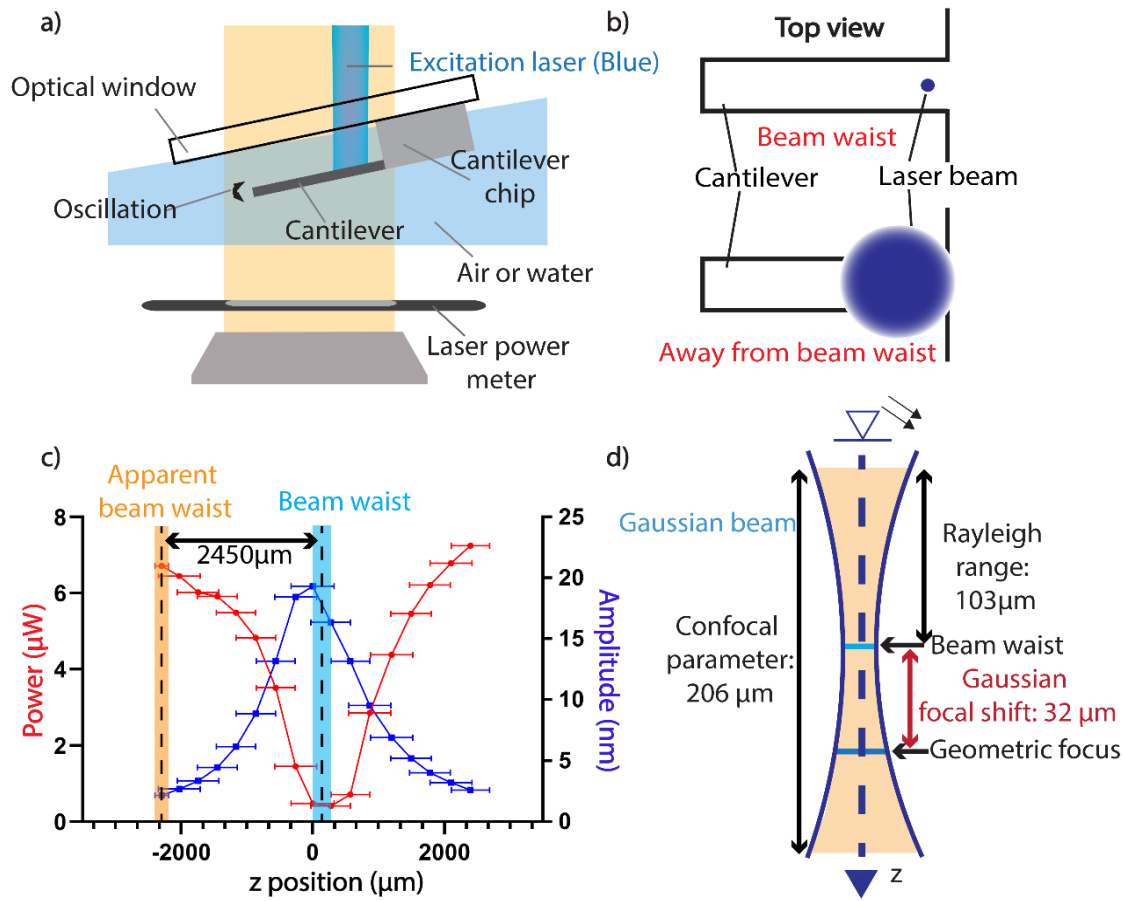

**Figure 3. Identifying the true position of the photothermal beam waist. a)** Schematic setup for measuring the residual optical power of the photothermal excitation beam beneath the cantilever in which the longitudinal position of the phothermal laser with respect to the cantilever is changed. To avoid interference, the infrared laser is switched off during these measurements. It is switched back on immidately afterwards to read out the cantilever oscillation amplitude. **b)** Top views schematic of the photothermal beam spot size at the cantilever for two different relative longitudinal cantilever positions. **c)** Cantilever (R500 TL) resonant oscillation amplitude and residual optical laser power versus longitudinal position measured using the setup in a). The apparent beam waist position is marked with a dashed line shaded orange and the real beam waist position is shown with a dash line shaded in blue. **d)** Schematic of the photothermal Gaussian beam profile with the Gaussian focal shift, Rayleigh range, confocal range and beam waist indicated.

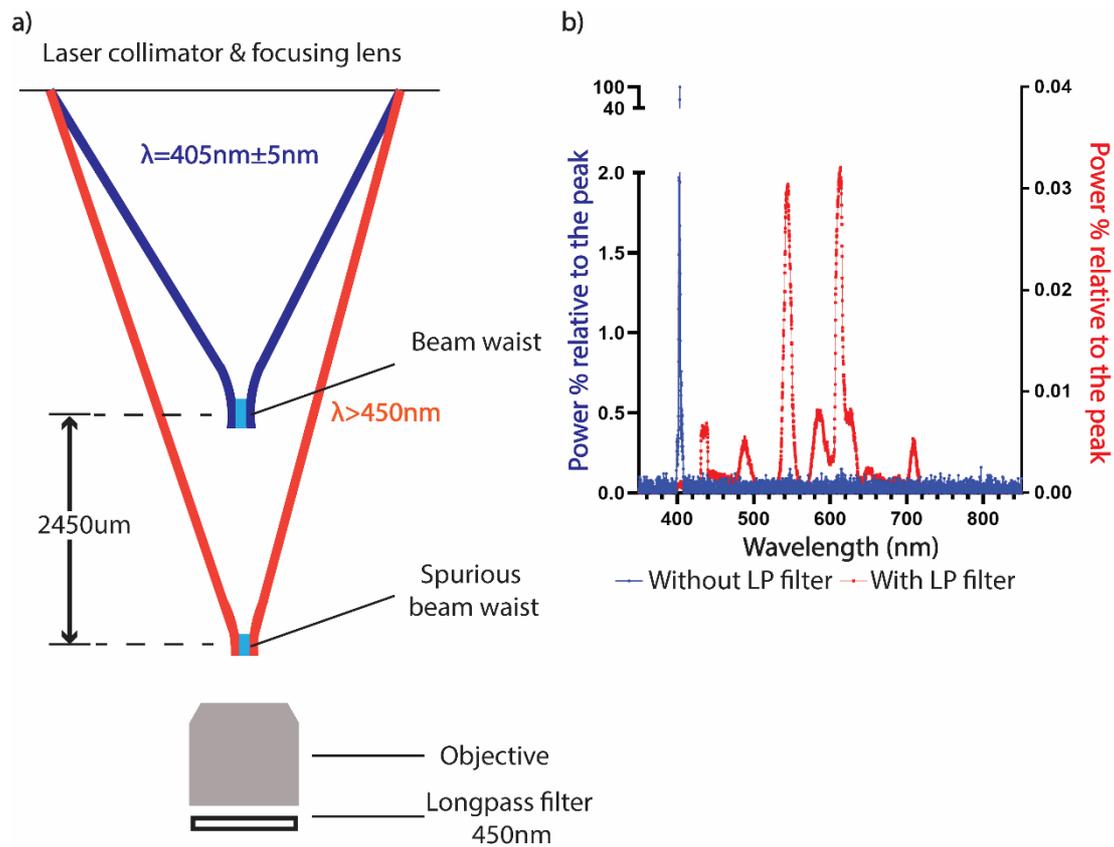

**Figure 4. Beams of different wavelengths generated by a laser diode. a)** Schematic of two superimposed beams of different wavelengths generated by a laser diode. The principal beam comprises the intended 405 nm light, yet the diode also generates light above 450 nm (spurious beam). The location of the beam waists of the principal and spurious beams differ by 2450 μm in our setup. **b)** Power spectra in the range 350–850 nm of the 405 nm laser diode use in the experiments without the 450 nm longpass (LP) filter (blue curve) and with the 450 nm LP filter (red curve). All measurements are normalised with respect to the peak laser power measured without LP filter.

**Table 1**: The detailed parameters of the cantilevers used in the experiment, including manufacturer, model, morphology, nominal dimensions, types of materials and coating layers. The oscillation information was obtained from the experiment.

| Manufacturer | Model | Beam number | Shape / Cross section | Dimensions (LxWxT) ($\mu m$) | Materials of cantilever | Backside coating | Oscillation in air | | | Oscillation in water | |
|---|---|---|---|---|---|---|---|---|---|---|---|
| | | | | | | | $f$ (kHz) | $Q$ | $k$ (N/m) | $f$ (kHz) | $Q$ |
| Nunano | QUEST R500 TL | N/A | Rectangular / Rectangular | 125x30x0.9 | Silicon nitride | 5nm/40 nm Ti/Au | 74.5 | 100 | 0.72 | 24.0 | 3.1 |
| Bruker | NP-010 | D | V-shaped / Rectangular | 200x25x0.55 | Silicon nitride | 5nm/45 nm Ti/Au | 19.1 | 32 | 0.083 | 4.11 | 1.7 |
| Bruker | NP-010 | C | V-shaped / Rectangular | 115x25x0.55 | Silicon nitride | 5nm/45 nm Ti/Au | 62.4 | 65 | 0.43 | 16.7 | 2.5 |
| MikroMasch | NSC35/TIPLESS/CR-AU | C | Rectangular / Trapezoidal | 145x35x2.0 | Silicon | 20nm/30 nm Cr/Au | 158 | 230 | 4.4 | 63.9 | 6.1 |
| MikroMasch | NSC35/TIPLESS/CR-AU | A | Rectangular / Trapezoidal | 126x35x2.0 | Silicon | 20nm/30 nm Cr/Au | 214 | 280 | 6.3 | 88.8 | 6.9 |
| MikroMasch | NSC35/TIPLESS/NO AL | A | Rectangular / Trapezoidal | 126x35x2.0 | Silicon | None | 216 | 320 | 7.4 | 83.9 | 6.4 |

# Supplementary information

Optimisation and artifacts of photothermal excitation of microresonators

*Liping Kevin Ge[1,2], Alessandro Tuniz[2,3], Martijn de Sterke[2,3], James M. Zavislan[4], Thomas G. Brown[4], Sascha Martin[5], and David Martinez-Martin[\*][1,2]*

[1] School of Biomedical Engineering, Faculty of Engineering and IT, University of Sydney, NSW, 2006 Australia
[2] Sydney Nano Institute. University of Sydney, NSW, 2006, Australia
[3] Institute of Photonics and Optical Science (IPOS), School of Physics, University of Sydney, Sydney, NSW, 2006, Australia
[4] The Institute of Optics, University of Rochester, Rochester, New York, 14627, USA
[5] Department of Physics, University of Basel, Basel, 4056, Switzerland

[\*] *Correspondence:* David Martinez-Martin: Email | david.martinezmartin@sydney.edu.au

Section 1

As shown in Figure S1, the optical microscope does not locate properly the position of the photothermal beam waist. Locating the cantilever at the photothermal beam waist identified with the optical microscope can reduce the photothermal efficiency almost up to an order of magnitude.

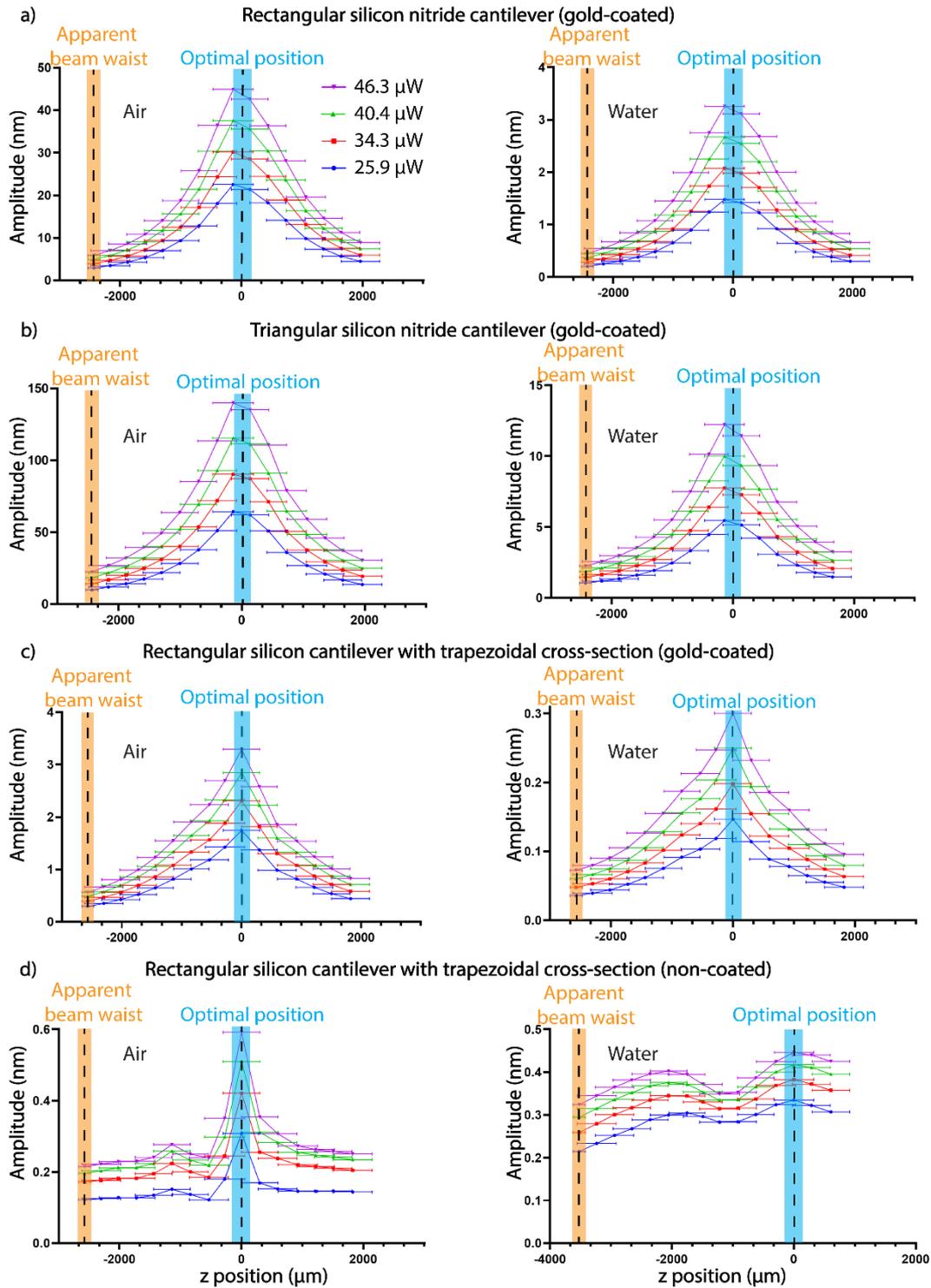

**Figure S1**. Cantilever amplitude at resonance for different positions of cantilevers along the propagation axis of the photothermal beam (z axis). The measurements are performed with different cantilever types in both air and water environments and for different optical powers of the photothermal beam. The origin of the z axis is chosen to be at the position of maximum photothermal efficiency, which is indicated by the dashed line labelled as "optimal position". The other dashed line, which is labelled as apparent beam waist, indicates the z position at which the cantilever is at the photothermal beam waist according to the optical microscope. **a)** Oscillation amplitudes at resonance for a Nunano R500 TL cantilever air (left graph) and water (right graph).

This is a gold coated silicon nitride cantilever with rectangular shape and rectangular cross section. **b)** Oscillation amplitudes at resonance for a Bruker NP-010D cantilever in air (left grapht) and water (right graph). This is a gold coated silicon nitride cantilever with triangular shape and rectangular cross section. **c)** Oscillation amplitudes at resonance for a MikroMasch NSC35/CR-AU-C cantilever in air (left graph) and water (right graph). This is a gold coated silicon cantilever with a rectangular shape and trapezoidal cross section. **d)** Oscillation amplitudes at resonance for a MikroMasch NSC35/NO AL-A cantilever in air (left graph) and water (right graph). This is an uncoated silicon cantilever with a rectangular shape and trapezoidal cross section. The initial starting point of the experiment along the z axis is at the apparent beam waist of the laser, and from there the cantilever is relatively moved along the z axis in increments of 300 μm. All results have been conducted for at least two independent cantilever probes and each probe has been repeated at least two times. For each laser position five consecutive resonance curves were acquired.

Section 2
We have characterised the chromatic aberration of the objective by using a pinhole. As shown in the figure S2 below we have illuminuated one side of a 2 µm diameter pinhole with light with two different wavelengths indenpendently, particularly using green light (550 nm wavelength) and our blue laser (405 nm wavelength). Upon illumination we observed the other side of the pinhole with the microscope and recorded the postion of the objective to get the pinhole in focus. The working distance of the objective varied slightly with the wavelength of the light used, with the objective working distance being approximately 50 µm larger for 405 nm wavelength light than that for 550 nm wavelength light. In this experiment the pinhole acted as light point source.

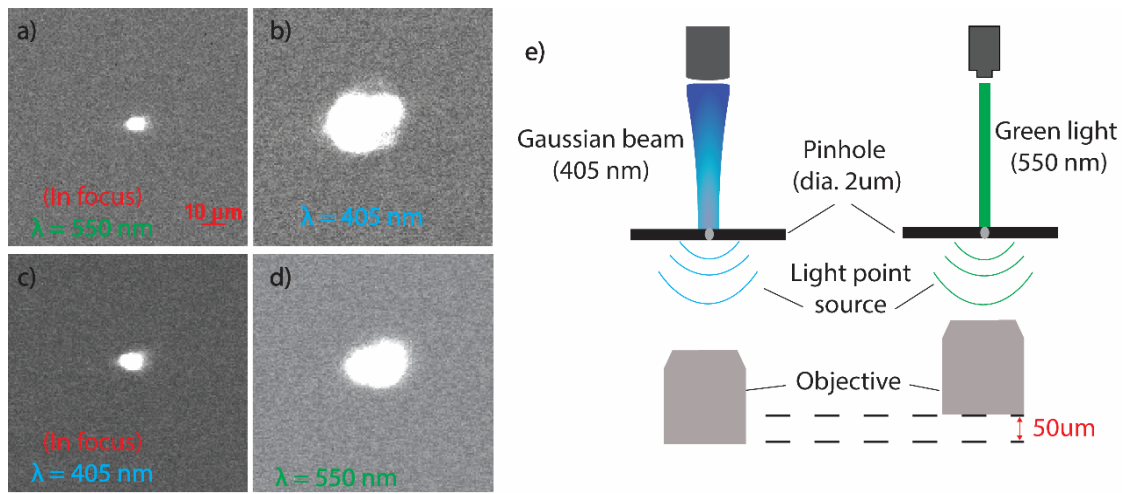

**Figure S2**. **Optical images of a 2µm-diameter pinhole illuminated with light of different wavelengths. a)** Image of the pinhole illuminated with 550 nm light. The image is in focus. **b)** Image of the pinhole illuminated by the 405 nm laser. The objective remains at the exact same distance from the pinhole as in a) and the pinhole appears now out of focus due to chromatic aberration. **c)** Image of the pinhole illuminated by the 405 nm laser as in b) after correcting the objective working distance, which had to be increased by 50 µm to get the pinhole in focus. **d)** Image of the pinhole illuminated by the 550 nm light with the objective at the same working distance used in c). The pinhole in these conditions appears out of focus. The scale bar is shown in a) applies to all the images. **e)** Schematic diagram of the experimental set up summarising the effect of the chromatic aberration.

Section 3

The laser diode that generates the 405 nm photothermal beam, generates additional beams with other wavelengths and foci than that of the intended 405 nm beam. Due to that effect the optical microscope mislocates the position of the photothermal beam. Such artifact can be eliminated by filtering the light generated by the diode with an appropriate band pass filter as confirmed below in the Figure S3.

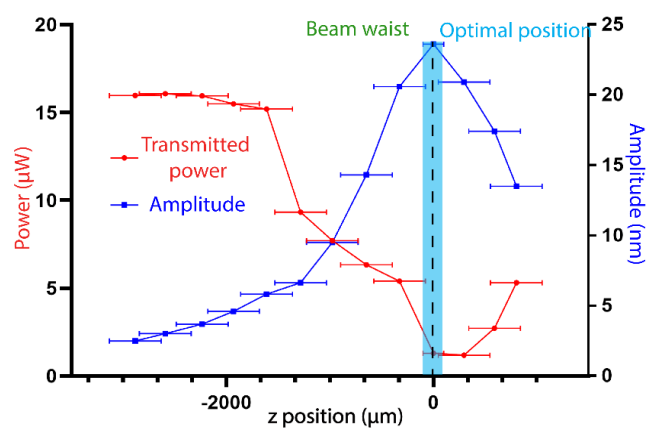

**Figure S3. Cantilever oscillation amplitude at resonance along the beam propagation axis (z axis) and residual (unscreened by the cantilever) optical laser power.** In this case the blue laser is filtered by a bandpass (BP) filter with centre wavelength of 405 nm and a bandwith of 10 nm. This filter removes the spurius light generated by the diode enabling to truly identify photothermal beam waist with the microscope. The beam waist identified with the microscope coincide with the real beam waist identified by the optical power measurements. The measurements are performed with a R500 TL cantilever in air. For this type of cantilever, the optimal excitation z position coincides approximately with the beam waist. The origin of the z axis is chosen to be at the position of maximum photothermal efficiency, which is indicated by the dash line lable as "optimal position".

Section 4

As demonstrated in Figure S3, using an appropriate bandpass filter for the photothermal beam enables to use the microscope to correctly locate the beam waist of the photothermal beam. For silicon nitride cantilevers with a rectangular cross section, the optimal position of the cantilever longitudinal position coincides approximately with the beam waist within the confocal parameter of the beam. That is indeed the expected since the maximum density of optical power is within that region. However, we find that for trapezoidal silicon cantilevers the optimal position for photothermal excitation is outside the confocal region of the beam.

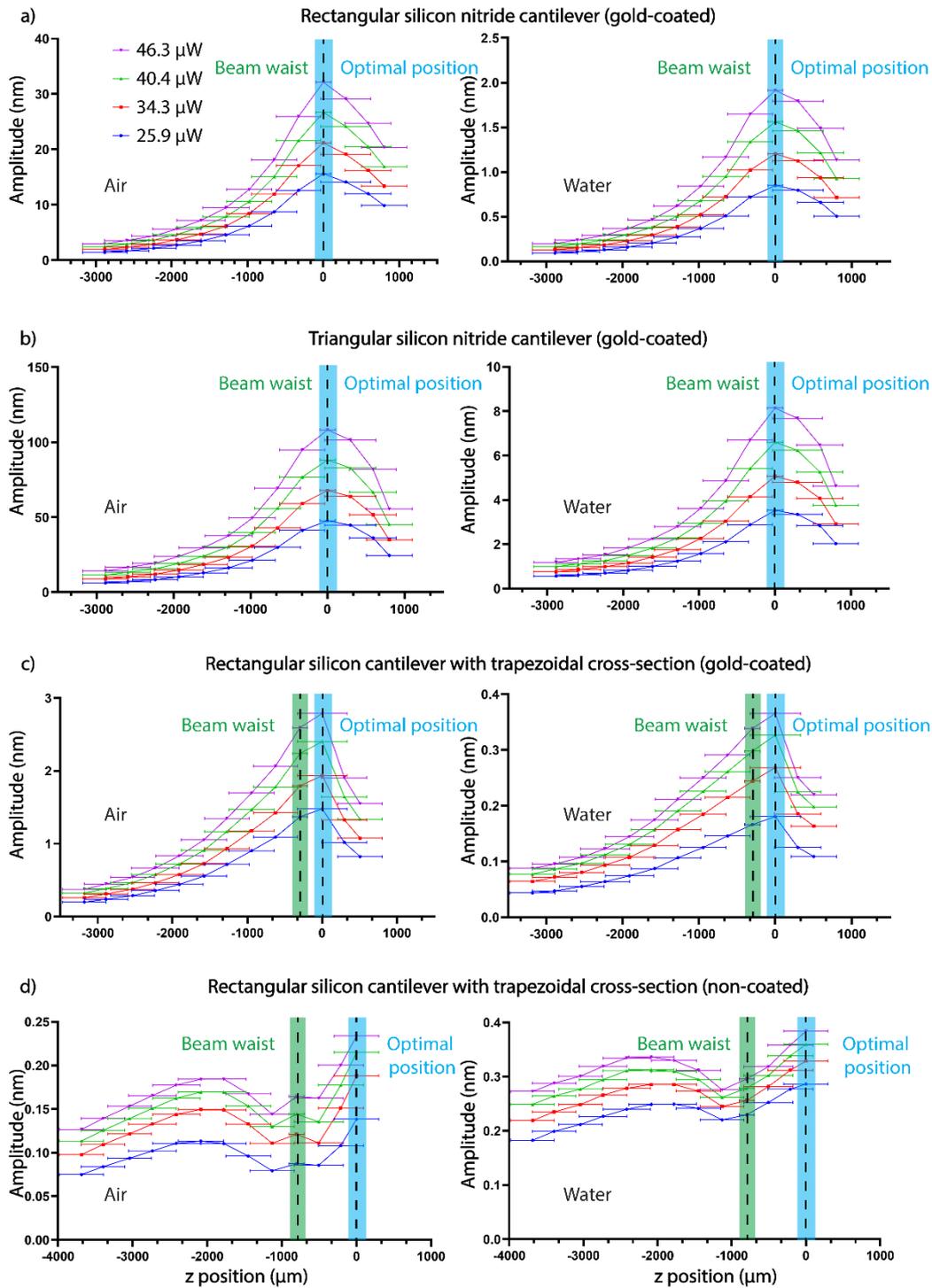

**Figure S4. Cantilever amplitude at resonance for different positions of cantilevers along the propagation axis of the photothermal beam (z axis).** This is a similar experiment as the one shown in Figure S2. However, in this case the experiment was performed filtering the photothermal laser with the BP filter with center wavelength of 405 nm and bandwith of 10 nm. As demonstrated in Figures S3, the BP filter removes spurious wavelengths generated by the diode that focus at a difference distance than the intended 405 nm radiation. The use of the BP filter enables to locate the true photothermal beam waist using the optical microscope. The green shadowed dashed line, labelled as "beam waist", indicates the longitudinal position at which the cantilever is at the beam waist of the filtered photothermal beam, as identified with the optical

microscope. The origin of the longitudinal position is chosen to be at the position of maximum photothermal efficiency, which is indicated by the dashed line labelled as "optimal position". **a)** Oscillation amplitudes at resonance for a Nunano R500 TL cantilever air (left graph) and water (right graph). This is a gold coated silicon nitride cantilever with rectangular shape and rectangular cross section. **b)** Oscillation amplitudes at resonance for a Bruker NP-010D cantilever in air (left graph) and water (right graph). This is a gold coated silicon nitride cantilever with triangular shape and rectangular cross section. **c)** Oscillation amplitudes at resonance for a MikroMasch NSC35/CR-AU-C cantilever in air (left graph) and water (right graph). This is a gold coated silicon cantilever with a rectangular shape and trapezoidal cross section. **d)** Oscillation amplitudes at resonance for a MikroMasch NSC35/NO AL-A cantilever in air (left graph) and water (right graph). This is a non-coated silicon cantilever with a rectangular shape and trapezoidal cross section. The optimal position could be further away, however, due to the space taken by the BP filter, there was not enough movement range to measure the further points for the uncoated silicon cantilevers. The initial starting point of the experiment is with the cantilever located at the beam waist. All results have been conducted for at least two independent cantilever probes and each probe has been repeated at least two times. For each laser position five consecutive resonance curves were acquired. The shape and the cross-section of each cantilever is labelled in the figure.